# Fermi Constraints on the Ejecta Speed and Prompt Emission Region of the Distant GRB 220101A

Lorenzo Scotton[1], Frédéric Piron[1], Nicola Omodei[2], Niccolò Di Lalla[2], and Elisabetta Bissaldi[3]
[1] Laboratoire Univers et Particules de Montpellier, CNRS/IN2P3, Place Eugène Bataillon—CC 72, Montpellier, France; lorenzoscotton@live.it
[2] Hansen Experimental Physics Lab, Stanford University, 452 Lomita Mall, Stanford, CA, USA
[3] Dipartimento di Fisica "M. Merlin" dell'Università e del Politecnico di Bari, via Amendola 173, I-70126 Bari, Italy
*Received 2023 May 10; revised 2023 August 9; accepted 2023 August 27; published 2023 October 12*

## Abstract

At redshift $z = 4.618$, GRB 220101A is the most distant gamma-ray burst (GRB) detected by Fermi/LAT to date. It is also a very energetic event, with an equivalent isotropic energy of $3.6 \times 10^{54}$ erg. We jointly analyzed the Fermi/GBM and LAT observations of GRB 220101A with two independent approaches and found a significant spectral break at sub-100 MeV energies during the prompt emission. The fast variability of the emission suggests that this spectral attenuation is caused by internal opacity to pair creation. Regardless of the nature of the emission processes assumed in the spectral analysis, we infer a moderate value for the jet Lorentz factor, $\Gamma \sim 110$, and find that all of the high-energy emission was produced above and near the photosphere, at a distance of $\sim 10^{14}$ cm from the central engine. We compare these results with the four other LAT-detected GRBs with similar properties.

*Unified Astronomy Thesaurus concepts:* Gamma-ray bursts (629); High energy astrophysics (739)

## 1. Introduction

Gamma-ray bursts (GRBs) are extragalactic and extremely energetic transient emissions of gamma rays. Their high luminosities suggest that the central engine of a GRB is a newborn stellar-mass black hole, which emits an ultrarelativistic collimated outflow (jet). At a typical distance from the central engine of $R \sim 10^{11}$–$10^{12}$ cm, the jet becomes transparent to thermal radiation, which is free to travel and possibly observed as a thermal component of the GRB spectrum. At an intermediate distance of $R \sim 10^{14}$–$10^{15}$ cm, still within the jet, either the kinetic energy carried by the jet dissipates via shocks or magnetic reconnection takes place. As a common result, charged particles are accelerated and emit highly variable synchrotron radiation. Both the thermal radiation, possibly reprocessed below the photosphere, and the nonthermal synchrotron radiation emitted at this intermediate region represent the prompt emission of the GRB. At larger radii, $R \sim 10^{16}$–$10^{17}$ cm, the jet collides with the circumburst medium, and the generated external shock accelerates charged particles that emit synchrotron radiation in this so-called afterglow phase. The prompt GRB emission is a short phase of intense and highly variable emission in hard X-rays and gamma rays that lasts from fractions of seconds to hundreds of seconds, while the subsequent afterglow phase is a long-lasting (hours, days) and decaying emission from (very) high energies (GeV–TeV) down to radio frequencies.

The first GRB catalog of the Burst and Transient Source Experiment on board the Compton Gamma Ray Observatory revealed a bimodality in the temporal and spectral distribution of GRBs (Kouveliotou et al. 1993); short GRBs have a duration of less than ∼2 s and are characterized by harder spectra, while long GRBs have a duration greater than ∼2 s and are typically softer. Short GRBs are believed to be produced by the merger of two neutron stars (Eichler et al. 1989; Narayan et al. 1992; Piran 2004) or a neutron star and a stellar-mass black hole (Paczynski 1991; Piran 2004). On 2017 August 17, the direct association of the gravitational wave GW 170817 emitted by the merger of a binary neutron star system and the short GRB 170817A (Abbott et al. 2017) proved that binary neutron star mergers are the progenitors of at least some short GRBs. On the other hand, long GRBs are believed to be produced by the collapse of fast-rotating massive stars (>30 $M_{Sun}$, Collapsar model; Woosley 1993; Piran 2004), as suggested by the association of nearby long GRBs with core-collapsed supernovae of Types Ib/Ic (Galama et al. 1998; Bloom et al. 2002; Hjorth et al. 2003; Piran 2004). In both scenarios, the merger of two compact objects or the collapse of a massive star result in the formation of a stellar-mass black hole, which acts as the central engine powering the jet.

The variable high-energy emission of some bursts, such as GRB 090926A (Yassine et al. 2017), GRB 100724B, GRB 160509A (Vianello et al. 2018), and GRB 170405A (Arimoto et al. 2020), exhibits a cutoff at the high end of its spectrum, which has been interpreted as a flux attenuation caused by the opacity to pair creation. In these rare cases, the theoretical framework developed by Hascoët et al. (2012) and applied by Yassine et al. (2017) on GRB 090926A allows one to directly determine the bulk Lorentz factor $\Gamma_{bulk}$ of the relativistic outflow and to localize the region where the observed variable high-energy emission was produced. This theoretical model assumes that the observed radiation is emitted close to or above the photosphere, and it does not rely on the specific nature of the emission mechanism but rather only on the knowledge of the burst distance, its emission variability, its broadband spectrum, and the cutoff energy.

The Fermi Gamma Ray Space Telescope is an observatory sensitive in the energy range from 10 keV to more than 300 GeV. It hosts two instruments: the Large Area Telescope (LAT; Atwood et al. 2009), which is an imaging, wide field-of-view (FOV), high-energy pair conversion telescope that covers the energy range from 20 MeV to more than 300 GeV, and the Gamma-ray Burst Monitor (GBM; Meegan et al. 2009), which comprises 12 sodium iodide (NaI) scintillation detectors and two bismuth germanate (BGO) detectors and covers the energy range from 8 keV to 40 MeV. The LAT standard analyses consider LAT data above 100 MeV and do not overlap with the energy range covered by the GBM, where the bulk of the GRB prompt emission is expected. Pelassa et al. (2010) proposed a nonstandard analysis technique to consider LAT data down to ∼20 MeV in order to fill this gap, thus providing useful data to better constrain the high-energy part of the GRB prompt spectra. These LAT low-energy (LLE) data are defined by less stringent cuts than LAT standard data, and they provide higher photon statistics above 100 MeV.

In this work, we analyze the exceptionally bright and distant GRB 220101A during its prompt emission at high energy using Fermi data, and we provide a physical interpretation of the observed emission. We specify the LAT and GBM data observations of GRB 220101A in Section 2.1, and we present the broadband spectral analysis procedure and results in Sections 2.2 and 3. Finally, we propose the interpretation of our results in Section 4 and compare GRB 220101A with other similar LAT-detected bursts in Section 5.

## 2. Observations and Analysis Procedure

### 2.1. Observations and Data Sets

The long GRB 220101A was detected and observed in a broad multiwavelength range. The prompt emission has been observed from hard X-rays to high-energy gamma rays, and the afterglow has been detected from optical (de Ugarte Postigo et al. 2022; Hentunen et al. 2022; Perley 2022) down to radio wavelengths up to few days after the event (Laskar 2022). The first detection of GRB 220101A was provided by the BAT instrument on board the Neil Gehrels Swift Observatory (Gehrels et al. 2004) at 05:10:12 UT on 2022 January 1 (first BAT notice).[4] This observatory also performed follow-up observations with XRT in the hard X-rays and UVOT in the visible domain (Tohuvavohu et al. 2022). Swift-UVOT localized GRB 220101A at R.A., decl. = 1°.35340, 31°.76903 with a 90% confidence error radius of 0″.61. Its photometric redshift was first measured by the Xinglong 2.16 m telescope at $z = 4.618$ (Fu et al. 2022) and later confirmed by the Liverpool Telescope (Perley 2022) and the Nordic Optical Telescope (Fynbo et al. 2022).

The Fermi/GBM triggered on GRB 220101A at $T_0 =$ 05:10:12 UT on 2022 January 1 (Lesage et al. 2022). The burst was also detected by Fermi/LAT at high energies (Arimoto et al. 2022) and occurred 18° from the LAT boresight at $T_0$. The LAT on-ground localization of the event is R.A., decl. = 1°.52, 31°.75 with an error radius of 0°.46, consistent with the Swift/XRT localization. The GBM data used in this work are the time-tagged events recorded by NaI detectors 3, 6, and 7, which observed the burst at an angle smaller than 60°, and by BGO detector 1, which was closest to the direction of the event at $T_0$. We also used the LAT standard P8R3_TRAN-SIENT020E_V2 data extracted from a region centered at the localization provided by the XRT with a 12° radius. Additionally, we used the LLE data to extend our analysis down to 20 MeV. Figure 1 shows the Fermi multidetector light curve of GRB 220101A during its prompt emission. The red dashed vertical line denotes the time of the trigger $T_0$, and the black dashed lines define the four time bins A, B, C, and D that are used in the time-resolved spectral analysis. The duration of the prompt emission is estimated as $T_{90} = (128 \pm 16)$ s (HEASARC GBM Burst Catalog).[5] The main emission episode in the GBM energy range (8 keV–40 MeV) was observed in the time interval $T_0 + [65, 134]$ s (time bins A–D), while the largest portion of LAT events is observed in the time interval $T_0 + [95, 107]$ s (time bins B and C). The brightest emission episode around $T_0 + 100$ s was jointly detected by the GBM detectors and the LAT. Interestingly, the high-energy flux is attenuated above ∼100 MeV during this episode. The highest-energy photon associated with the burst with a probability greater than 99% was detected at a later time ($T_0 + 152$ s) with an energy of 927 MeV. In this work, we focus on the brightest emission episode around $T_0 + 100$ s. The variability of this emission as seen in Figure 1 suggests that it has an internal origin in the jet. Consistently, Mei et al. (2022) interpreted this episode as prompt-dominated, with an afterglow appearing only after ∼118 s.

### 2.2. Analysis Procedure

First, we performed an LAT-only standard analysis based on the unbinned likelihood method using `fermitools`.[6] We employed the likelihood ratio test (LRT; Neyman & Pearson 1928) to estimate the significance of the GRB detection with the LAT. In the null hypothesis, the background model is composed of the isotropic emission only, which is typically fitted as a power-law (PL) spectrum. The contribution to the background in the LAT from the galactic diffuse emission was neglected owing to the high latitude of the burst (∼−30°). A detection threshold TS$_{\rm GRB}$ > 20 was then used following the first LAT GRB catalog (Ackermann et al. 2013), which corresponds to a one-sided Gaussian probability of 4.1σ. We also used the LRT to search for spectral attenuation at high energies using an exponential cutoff multiplicative model. The corresponding test statistic is defined as TS$_{\rm cut}$ = $-2 \ln[\mathcal{L}_{\rm max}(M_0)/\mathcal{L}_{\rm max}(M_1)]$, where $M_0$ is the spectral model in the null hypothesis, $M_1 = M_0 \times \exp(-E/E_{\rm cut})$ is the spectral model in the alternate hypothesis, and $E_{\rm cut}$ is the cutoff energy. In the LAT-only standard analysis, $M_0$ is a PL, and $M_1$ is referred to as CUTPL. Since TS$_{\rm cut}$ follows a $\chi^2$ with one degree of freedom (the additional $E_{\rm cut}$ parameter) in the large sample limit (Wilks 1938), we estimated the Gaussian significance of the additional cutoff as $\sigma_{\rm cut} = \sqrt{\rm TS_{cut}}$.

Next, we performed a joint GBM–LAT spectral analysis. We used the `gtburst` software to bin the event data and produce a total count spectrum in each GBM detector. This tool was also used to create GBM background count spectra during the GRB on-source interval by fitting polynomial functions in each energy channel of two off-source intervals and extrapolating them to the GRB interval. In addition, we used `fermitools` to create the LAT count spectra from the best-fit model obtained in the LAT-only standard analysis. We jointly analyzed the GBM and LAT count spectra with the `pyXSPEC` fitting software[7] (Arnaud 1996). In order to check the stability of our results, we also performed a joint spectral analysis using the "Multi-Mission Maximum Likelihood" (`threeML`) software[8] (Vianello et al. 2015), which allows us to combine the native likelihoods of different instruments simultaneously. In the current analysis, `threeML` offered the full accuracy of the LAT unbinned likelihood technique, which is lost during the binning in space and energy that is required by `pyXSPEC`. We considered the following spectral models, which are differential photon energy spectra in units of $cm^{-2}\,s^{-1}\,keV^{-1}$.

---

[4] https://heasarc.gsfc.nasa.gov/wsgi-scripts/tach/gcn_v2/tach.wsgi/
[5] https://heasarc.gsfc.nasa.gov/W3Browse/fermi/fermigbrst.html
[6] https://fermi.gsfc.nasa.gov/ssc/data/analysis/documentation/Cicerone/Cicerone_Likelihood
[7] https://heasarc.gsfc.nasa.gov/xanadu/xspec/python/html/index.html
[8] https://threeml.readthedocs.io/en/latest/index.html

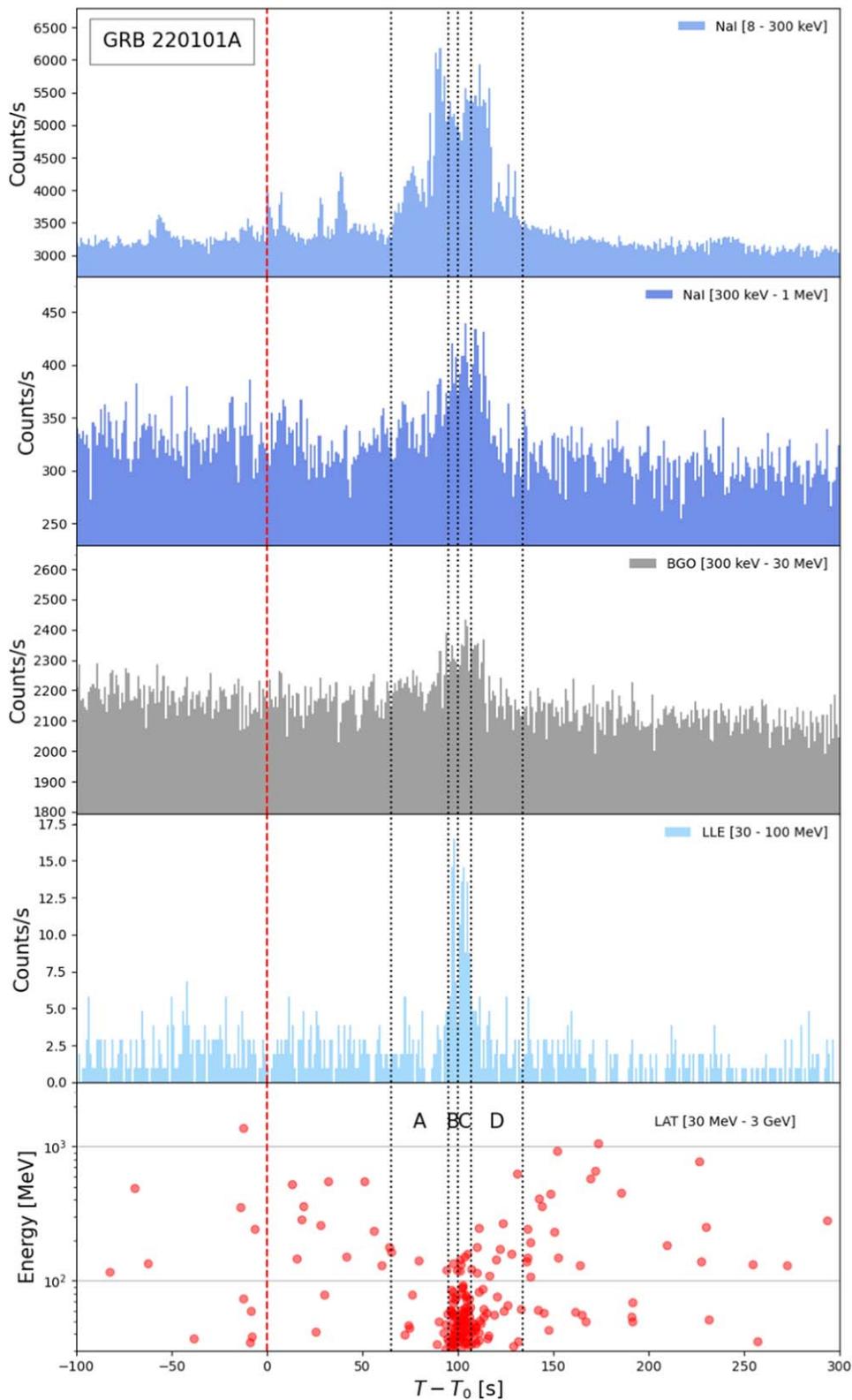

**Figure 1.** Fermi multidetector light curve of GRB 220101A prompt emission in increasing energy bands from top to bottom. The first four panels present count rates, while the last panel presents the energy of the LAT-observed events. The red dashed vertical line denotes the time of the trigger $T_0$, while the black dashed vertical lines indicate the time intervals chosen for the time-resolved spectral analysis, covering the main emission episode observed by the LAT.

**Table 1**
Results of the LAT-only Spectral Analysis of PL and CUTPL in Different Time Windows

| $T - T_0$ (s) | Range (MeV) | PL | | CUTPL | | | |
|---|---|---|---|---|---|---|---|
| | | Index | TS | Index | $E_{\text{cut}}$ (MeV) | TS | $\sigma_{\text{cut}}$ |
| 0–600 | >100 | $-2.48 \pm 0.23$ | **104.1** | $-1.97 \pm 0.58$ | $939 \pm 1129$ | **105.3** | 1.1 |
|  | >30  | $-2.93 \pm 0.13$ | **170.0** | $-2.93 \pm 0.13$ | $(2.9 \pm 7.9) \times 10^5$ | **170.0** | 0 |
| 0–65 | >100 | $-2.33 \pm 0.74$ | 10.6 | $-1.01 \pm 0.37$ | $321 \pm 321$ | 1.1 | 0.7 |
|  | >30 | $-1.73 \pm 0.39$ | 12.7 | $-1.00 \pm 0.02$ | $458 \pm 434$ | 14.6 | 1.4 |
| 65–134 | >100 | $-3.41 \pm 0.52$ | **45.7** | $-2.97 \pm 1.32$ | $607 \pm 1816$ | **45.8** | 0.3 |
|  | >30 | $-3.48 \pm 0.17$ | **129.1** | $-3.45 \pm 0.28$ | $3167 \pm 21570$ | **129.1** | 0 |
| 134–300 | >100 | $-2.18 \pm 0.31$ | **47.3** | $-1.00 \pm 0.08$ | $427 \pm 193$ | **50.0** | 1.6 |
|  | >30 | $-1.98 \pm 0.21$ | **56.8** | $-1.0 \pm 2.3$ | $439 \pm 1568$ | **60.7** | 2.0 |
| 300–600 | >100 | $-1.81 \pm 0.51$ | 11.1 | $-1.00 \pm 0.08$ | $945 \pm 931$ | 12.1 | 1.0 |
|  | >30 | $-1.76 \pm 0.50$ | 9.9 | $-1.00 \pm 0.01$ | $1045 \pm 1143$ | 10.7 | 0.9 |

**Notes.** The units of the normalization are $10^{-7}\,\text{cm}^{-2}\,\text{s}^{-1}\,\text{MeV}^{-1}$. The bold values correspond to the cases where the spectral model in the alternate hypothesis is preferred to the spectral model in the null hypothesis, as explained in Section 2.2.

1. Band (four parameters). Introduced by Band et al. (1993), it reads

$$f_{\text{Band}}(E) = A \times \begin{cases} \left(\dfrac{E}{E_{\text{piv}}}\right)^\alpha \exp\left[-\dfrac{E(2+\alpha)}{E_p}\right] & \text{if } E \leqslant E_b = E_p \dfrac{\alpha-\beta}{2+\alpha} \\ \left(\dfrac{E}{E_{\text{piv}}}\right)^\beta \exp(\beta-\alpha)\left[\dfrac{E_p(\alpha-\beta)}{E_{\text{piv}}(2+\alpha)}\right]^{\alpha-\beta} & \text{otherwise} \end{cases}$$

(1)

where $\alpha$ is the low-energy spectral index, $\beta$ is the high-energy spectral index, $E_p$ is the peak energy of the spectral energy distribution (SED), $E_b$ is the break energy, and $E_{\text{piv}}$ is the reference energy fixed to 100 keV.

2. Internal shock synchrotron model (ISSM; four parameters). It was introduced by Yassine et al. (2020) and further investigated by L. Scotton et al. (2023, in preparation) as a proxy function of the GRB internal shock model developed by Bošnjak & Daigne (2014),

$$f_{\text{ISSM}}(E) = \dfrac{A}{\left[1 - \dfrac{E_p}{E_r}\left(\dfrac{2+\beta}{2+\alpha}\right)\right]^{\beta-\alpha}} \times \left(\dfrac{E}{E_r}\right)^\alpha \left[\dfrac{E}{E_r} - \dfrac{E_p}{E_r}\left(\dfrac{2+\beta}{2+\alpha}\right)\right]^{\beta-\alpha},$$

(2)

where $\alpha$, $\beta$, and $E_p$ have the same meaning as in the Band model, and $E_r$ is the reference energy fixed to 10 keV.

We implemented both functions as local models in pyXSPEC and threeML. We also considered the models obtained by multiplying these functions by an exponential cutoff at high energies ($\propto e^{-E/E_{\text{cut}}}$). We called the resulting models BandExpCut and ISSMExpCut, respectively. Similar to the LAT-only standard analysis, we estimated the spectral cutoff significance as $\sigma_{\text{cut}} = \sqrt{\text{TS}_{\text{cut}}}$ using Band or ISSM for the $M_0$ spectral model in the null hypothesis and BandExpCut or ISSMExpCut for the $M_1$ spectral model in the alternate hypothesis.

## 3. Broadband Spectral Analysis Results

### 3.1. High-energy Spectral Evolution

We analyzed LAT standard data at energies greater than 100 MeV to search for a GRB detection and test whether a spectral cutoff is statistically required. We considered $T_0 + [0, 600]$ s as the time interval in which the burst position was in the LAT FOV. In particular, we focused on the main emission interval $T_0 + [65, 134]$ s and the time intervals $T_0 + [0, 65]$, [134, 300], and [300, 600] s.

Table 1 presents the analysis results. High-energy emission from the point source is detected (TS > 20) over the whole time interval $T_0 + [0, 600]$ s and, more specifically, in the main emission episode $T_0 + [65, 134]$ s and in $T_0 + [134, 300]$ s. No high-energy emission is detected before $T_0 + 65$ s and after $T_0 + 300$ s or in the time window when the burst reentered the LAT FOV, i.e., $4500\,\text{s} < T - T_0 < 6000\,\text{s}$. In the main emission interval $T_0 + [65, 134]$ s, its spectral index is very steep and significantly softer than $-3$. This is consistent with the depleted flux seen at $\sim T_0 + 100$ s in Figure 1. However, no cutoff is required by the data in any time interval. We thus increased the spectral coverage and sensitivity to a possible cutoff by including LAT data down to 30 MeV, ignoring the energy dispersion effects that are not implemented in the unbinned likelihood analysis. As expected, the PL index was better constrained, but no spectral cutoff was detected.

### 3.2. Time-resolved Prompt Emission Spectra

Since no spectral cutoff was detected in the LAT-only spectral analysis, we extended the energy lever arm to lower energy by including GBM data in the fits. Table 2 shows the best-fit parameters, fit statistics, and significance of the additional spectral cutoff of BandExpCut on GBM+LAT data in the four time intervals A, B, C, and D. The $E_{\text{cut}}$ is significantly detected in time bins B and C at $26 \pm 13$ and $45 \pm 13$ MeV, respectively. Moreover, we considered LLE data down to 20 MeV instead of the LAT standard data to properly

Table 2
pyXSPEC Spectral Fits of Band with and without a Cutoff on GBM+LAT Data in Time Intervals A, B, C, and D

| Parameter | A: $T_0 + [65, 95]$ s | | B: $T_0 + [95, 100]$ s | |
|---|---|---|---|---|
| | Band | BandExpCut | Band | BandExpCut |
| $\alpha$ | $-0.77 \pm 0.03$ | $-0.73 \pm 0.05$ | $-0.85 \pm 0.05$ | $-0.82 \pm 0.06$ |
| $\beta$ | $-2.71 \pm 0.06$ | $-2.21 \pm 0.33$ | $-2.46 \pm 0.05$ | $-1.9 \pm 0.2$ |
| $E_p$ [keV] | $248 \pm 12$ | $230 \pm 23$ | $384 \pm 39$ | $339 \pm 46$ |
| $E_{cut}$ [MeV] | … | $24 \pm 23$ | … | **$26 \pm 13$** |
| Norm. ($10^{-2}$) | $2.09 \pm 0.10$ | $2.22 \pm 0.19$ | $2.62 \pm 0.17$ | $2.76 \pm 0.22$ |
| PGSTAT/dof | 638/524 | 627/523 | 593/524 | 569/523 |
| $\sigma_{cut}$ | … | 3.3 | … | **4.9** |
| | C: $T_0 + [100, 107]$ s | | D: $T_0 + [107, 134]$ s | |
| | Band | BandExpCut | Band | BandExpCut |
| $\alpha$ | $-0.85 \pm 0.04$ | $-0.82 \pm 0.05$ | $-0.92 \pm 0.03$ | $-0.91 \pm 0.01$ |
| $\beta$ | $-2.41 \pm 0.03$ | $-2.04 \pm 0.09$ | $-2.56 \pm 0.04$ | $-2.4 \pm 0.4$ |
| $E_p$ [keV] | $441 \pm 38$ | $393 \pm 40$ | $299 \pm 17$ | $295 \pm 4$ |
| $E_{cut}$ [MeV] | … | **$45 \pm 13$** | … | $100 \pm 1$ |
| Norm. ($10^{-2}$) | $2.58 \pm 0.12$ | $2.72 \pm 0.16$ | $1.81 \pm 0.08$ | $1.82 \pm 0.01$ |
| PGSTAT/dof | 544/524 | 515/523 | 576/524 | 587/523 |
| $\sigma_{cut}$ | … | **5.4** | … | 0 |

**Notes.** The units of the normalization are cm$^{-2}$ s$^{-1}$ keV$^{-1}$. The bold values correspond to the cases where the spectral model in the alternate hypothesis is preferred to the spectral model in the null hypothesis, as explained in Section 2.2.

Table 3
pyXSPEC Spectral Fits of Band with and without the Cutoff on GBM+LLE Data in Time Intervals B, C, and B + C

| Parameter | B: $T_0 + [95, 100]$ s | | C: $T_0 + [100, 107]$ s | | B + C: $T_0 + [95, 107]$ s | |
|---|---|---|---|---|---|---|
| | Band | BandExpCut | Band | BandExpCut | Band | BandExpCut |
| $\alpha$ | $-0.85 \pm 0.05$ | $-0.82 \pm 0.05$ | $-0.85 \pm 0.04$ | $-0.82 \pm 0.04$ | $-0.85 \pm 0.03$ | $-0.82 \pm 0.03$ |
| $\beta$ | $-2.32 \pm 0.04$ | $-1.87 \pm 0.11$ | $-2.31 \pm 0.03$ | $-2.09 \pm 0.05$ | $-2.31 \pm 0.02$ | $-2.01 \pm 0.06$ |
| $E_p$ [keV] | $387 \pm 41$ | $337 \pm 39$ | $437 \pm 39$ | $397 \pm 35$ | $416 \pm 29$ | $373 \pm 26$ |
| $E_{cut}$ [MeV] | … | **$23 \pm 8$** | … | **$69 \pm 20$** | … | **$42 \pm 11$** |
| Norm. ($10^{-2}$) | $2.60 \pm 0.17$ | $2.8 \pm 0.2$ | $2.59 \pm 0.13$ | $2.70 \pm 0.14$ | $2.59 \pm 0.10$ | $2.73 \pm 0.12$ |
| PGSTAT/dof | 609/519 | 570/518 | 536/519 | 510/518 | 672/519 | 612/518 |
| $\sigma_{cut}$ | … | **6.2** | … | **5.1** | … | **7.7** |

**Notes.** The units of the normalization are cm$^{-2}$ s$^{-1}$ keV$^{-1}$. The bold values correspond to the cases where the spectral model in the alternate hypothesis is preferred to the spectral model in the null hypothesis, as explained in Section 2.2.

Table 4
pyXSPEC Spectral Fits of ISSM with and without a Cutoff on GBM+LLE Data in Time Intervals B, C, and B + C

| Parameter | B: $T_0 + [95, 100]$ s | | C: $T_0 + [100, 107]$ s | | B + C: $T_0 + [95, 107]$ s | |
|---|---|---|---|---|---|---|
| | ISSM | ISSMExpCut | ISSM | ISSMExpCut | ISSM | ISSMExpCut |
| $\alpha$ | $-0.77 \pm 0.06$ | $-0.66 \pm 0.09$ | $-0.74 \pm 0.07$ | $-0.67 \pm 0.10$ | $-0.75 \pm 0.05$ | $-0.67 \pm 0.09$ |
| $\beta$ | $-2.52 \pm 0.06$ | $-2.17 \pm 0.05$ | $-2.50 \pm 0.05$ | $-2.28 \pm 0.05$ | $-2.50 \pm 0.03$ | $-2.24 \pm 0.07$ |
| $E_p$ [keV] | $701 \pm 60$ | $1263 \pm 332$ | $776 \pm 69$ | $996 \pm 130$ | $751 \pm 46$ | $1066 \pm 236$ |
| $E_{cut}$ [MeV] | … | **$41 \pm 10$** | … | $88 \pm 27$ | … | **$64 \pm 22$** |
| Norm. ($10^{-2}$) | $18 \pm 1$ | $17 \pm 1$ | $17 \pm 1$ | $16 \pm 2$ | $17.2 \pm 0.9$ | $16 \pm 1$ |
| PGSTAT/dof | 605/519 | 582/518 | 530/519 | 517/518 | 661/519 | 629/518 |
| $\sigma_{cut}$ | … | **4.8** | … | 3.6 | … | **5.7** |

**Notes.** The units of the normalization are cm$^{-2}$ s$^{-1}$ keV$^{-1}$. The bold values correspond to the cases where the spectral model in the alternate hypothesis is preferred to the spectral model in the null hypothesis, as explained in Section 2.2.

account for the energy dispersion and benefit from the greater photon statistics. Table 3 shows the results of the Band fits with and without the spectral cutoff to time bins B, C, and B + C. We further checked that the results do not depend strongly on the specific choice of the Band model and also used the ISSM model to describe the nonthermal spectrum. Table 4 shows the corresponding results for ISSM, and Table 5 summarizes the overall results.

A spectral cutoff is detected in time bins B and B + C for both BandExpCut and ISSMExpCut, while it is detected in time bin C only when considering BandExpCut. We note that the significance of the spectral cutoff is systematically smaller when employing ISSMExpCut; the continuous curvature of ISSM, which reflects the natural shape of GRB synchrotron spectra, accounts for part of the softening of the spectra at high energies and thus reduces the significance of the additional cutoff. Figure 2 shows the GRB count spectra and residuals (upper panels) and SEDs (lower panels) when fitting Band (left panels) and BandExpCut (right panels) in time bin B + C. Figure 3 shows the same quantities for ISSM and ISSMExpCut. The residuals in the LLE energy range improve when adding the high-energy spectral cutoff to both Band and ISSM, and this is consistent with the significant detection of the spectral cutoff.

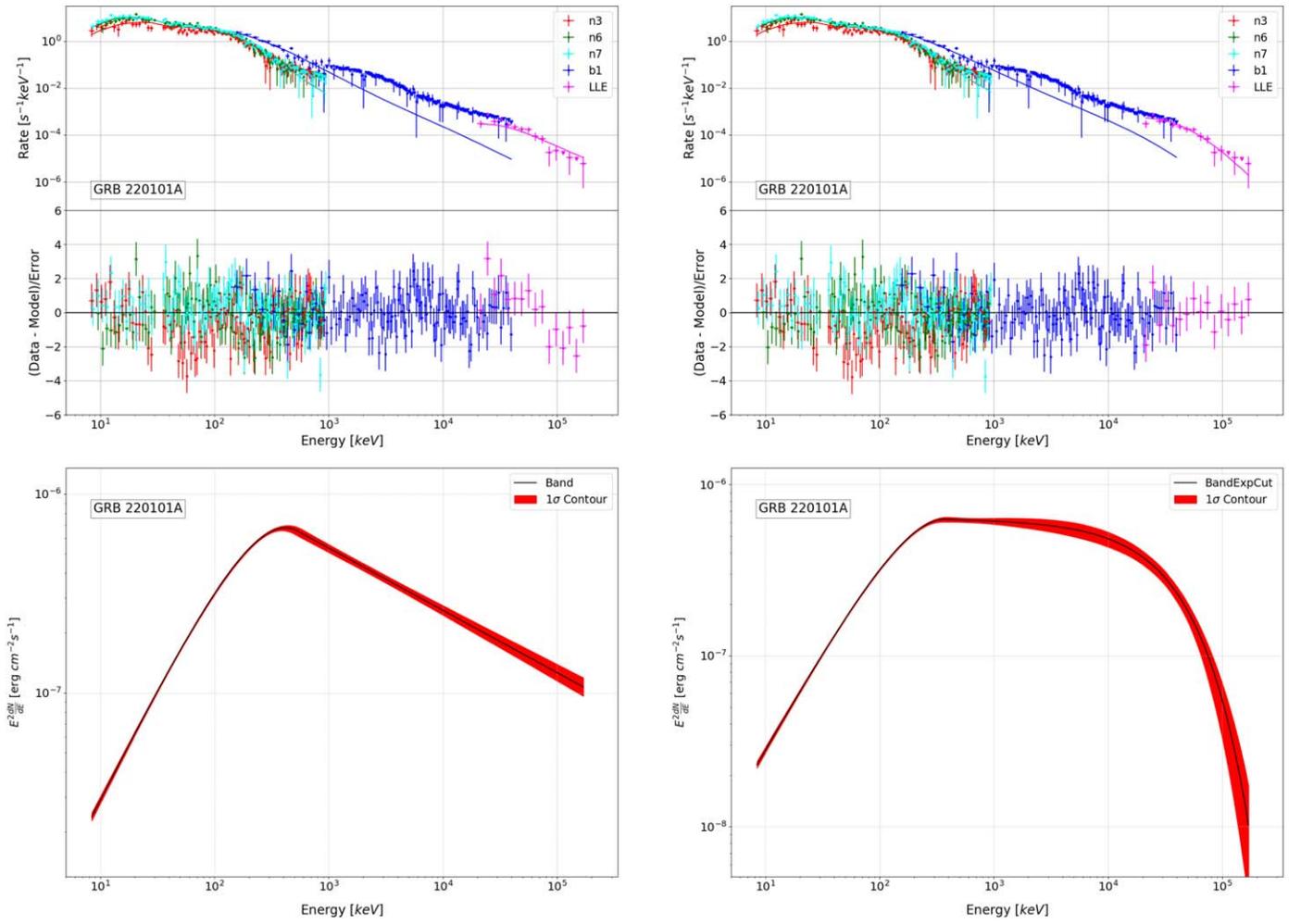

**Figure 2.** Left: GRB count spectra and residuals (upper panel) and SED (lower panel) from Band fits to GBM+LLE data in time bin B + C with `pyXSPEC`. Right: same for BandExpCut.

Table 5
Summary of the `pyXSPEC` Spectral Analysis

|  |  | B: $T_0 + [95, 100]$ s | C: $T_0 + [100, 107]$ s | B + C: $T_0 + [95, 107]$ s |
|---|---|---|---|---|
| BandExpCut | $E_{\rm cut}$ [MeV] | $23 \pm 8$ | $69 \pm 20$ | $42 \pm 11$ |
|  | $\sigma_{\rm cut}$ | 6.2 | 5.1 | 7.7 |
| ISSMExpCut | $E_{\rm cut}$ [MeV] | $41 \pm 10$ | $88 \pm 27$ | $64 \pm 22$ |
|  | $\sigma_{\rm cut}$ | 4.8 | 3.6 | 5.7 |

In order to assess possible systematic effects due to the specific software used for the spectral data preparation and fit,

we performed the same analysis within the framework of `threeML`. Tables 6 and 7 show the `threeML` spectral results when fitting Band and ISSM, respectively, with and without the spectral cutoff to GBM+LLE data. The spectral results are fully consistent between the two different approaches; the high-energy cutoff is required in time bins B ($5.8\sigma$), C ($4.7\sigma$), and B + C ($7.1\sigma$) when fitting BandExpCut, while it is required in time bins B ($4.7\sigma$), C ($3.2\sigma$), and B + C ($5.3\sigma$) when fitting ISSMExpCut. The results from `pyXSPEC` and `threeML` are in excellent agreement, confirming the cutoff detection already found with `pyXSPEC`.

As mentioned in Section 2.2, `threeML` makes use of the native LAT likelihood; therefore, we performed the same fits on GBM+LLE+LAT data, limiting the LLE data below 100 MeV and considering the LAT standard data above 100 MeV. The corresponding results are reported in Tables 8

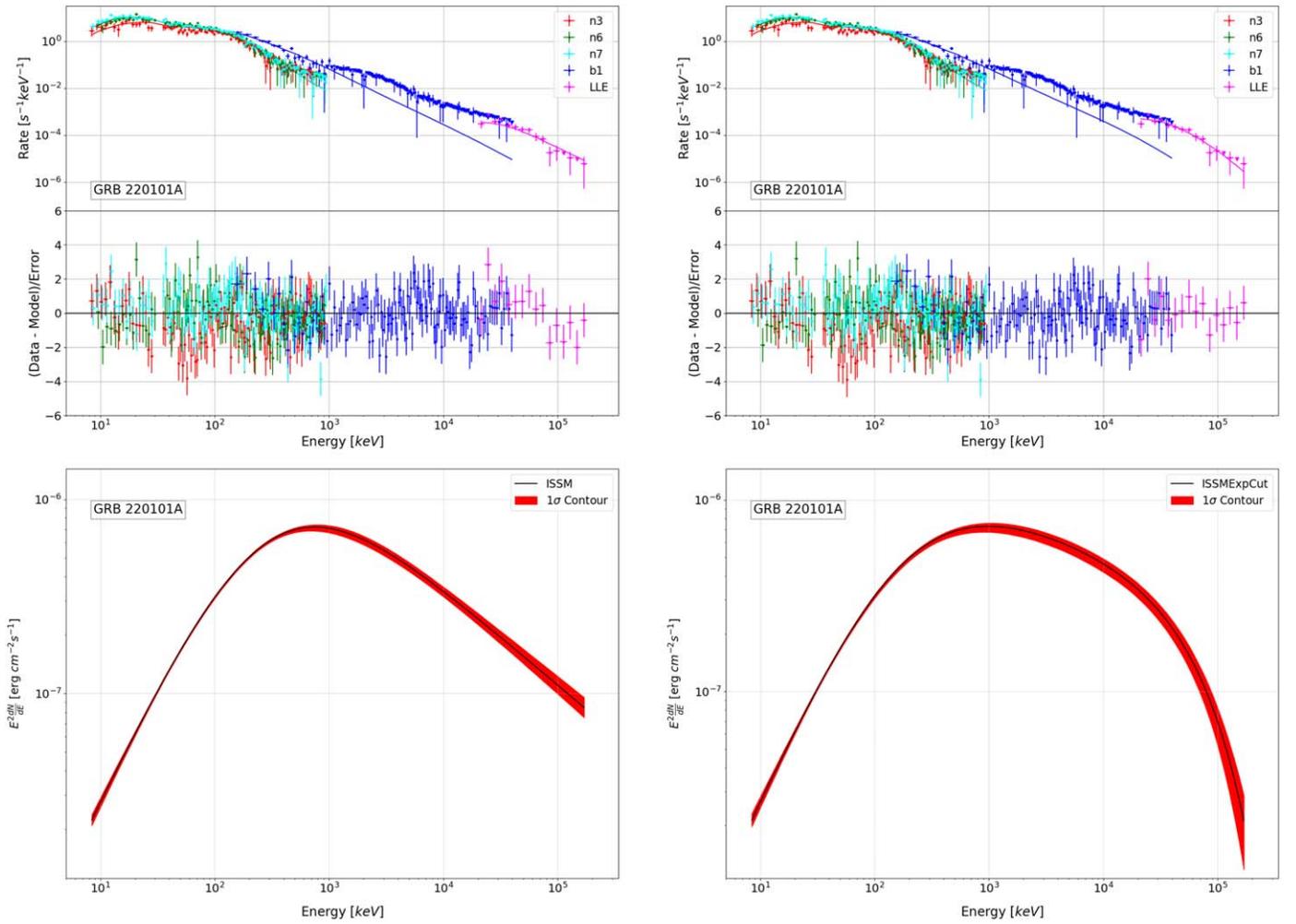

**Figure 3.** Left: GRB count spectra and residuals (upper panel) and SED (lower panel) from ISSM fits to GBM+LLE data in time bin B + C with `pyXSPEC`. Right: same for ISSMExpCut.

Table 6
`threeML` Spectral Fits of Band with and without the Cutoff on GBM+LLE Data in Time Intervals B, C, and B + C

| Parameter | B: $T_0 + [95, 100]$ s | | C: $T_0 + [100, 107]$ s | | B + C: $T_0 + [95, 107]$ s | |
|---|---|---|---|---|---|---|
| | Band | BandExpCut | Band | BandExpCut | Band | BandExpCut |
| $\alpha$ | $-0.84 \pm 0.05$ | $-0.80 \pm 0.06$ | $-0.85 \pm 0.04$ | $-0.82 \pm 0.04$ | $-0.84 \pm 0.03$ | $-0.81 \pm 0.04$ |
| $\beta$ | $-2.31 \pm 0.04$ | $-1.88 \pm 0.08$ | $-2.29 \pm 0.03$ | $-2.09 \pm 0.06$ | $-2.30 \pm 0.02$ | $-2.01 \pm 0.05$ |
| $E_p$ [keV] | $370 \pm 40$ | $320 \pm 40$ | $420 \pm 40$ | $390 \pm 40$ | $401 \pm 27$ | $359 \pm 26$ |
| $E_{\rm cut}$ [MeV] | ⋯ | **$23 \pm 6$** | ⋯ | **$71 \pm 25$** | ⋯ | **$43 \pm 11$** |
| Norm. ($10^{-2}$) | $2.60 \pm 0.17$ | $2.77 \pm 0.22$ | $2.60 \pm 0.13$ | $2.71 \pm 0.15$ | $2.60 \pm 0.10$ | $2.74 \pm 0.13$ |
| $-\log(\mathcal{L})$ | 2129 | 2112 | 2352 | 2341 | 2792 | 2767 |
| $\sigma_{\rm cut}$ | ⋯ | **5.8** | ⋯ | **4.7** | ⋯ | **7.1** |

**Notes.** The units of the normalization are cm$^{-2}$ s$^{-1}$ keV$^{-1}$. The bold values correspond to the cases where the spectral model in the alternate hypothesis is preferred to the spectral model in the null hypothesis, as explained in Section 2.2.

and 9. The likelihoods of the Band and ISSM fits are remarkably similar within the analyses of both GBM+LLE and GBM+LLE+LAT data. Table 10 resumes the results of the threeML spectral analysis: a spectral cutoff is detected at $32 \pm 9$ (6.2$\sigma$), $67 \pm 16$ (5.8$\sigma$), and $51 \pm 10$ (8.4$\sigma$) MeV when fitting BandExpCut in time bins B, C, and B + C, respectively. A high-energy cut is detected at $41 \pm 13$ (4.9$\sigma$), $88 \pm 28$ (4.5$\sigma$), and $66 \pm 15$ (6.3$\sigma$) MeV when fitting ISSMExpCut in the same time bins. As already observed, the significances of the spectral cutoffs in the case of ISSMExpCut are smaller due to the continuous curvature of ISSM. Moreover, the cutoff significance is larger than in the previous GBM+LLE data analysis, especially in time bins C and B + C. This can be explained by the better sensitivity of the native LAT likelihood, which manifests particularly in these time bins where the spectral cutoff is close to 100 MeV, the lower bound of LAT standard data.

The fits of BandExpCut and ISSMExpCut yield similar results within the errors but with different fitted spectral cutoff values. We used these differences to assess the systematic

Table 7
threeML Spectral Fits of ISSM with and without the Cutoff on GBM+LLE Data in Time Intervals B, C, and B + C

| Parameter | B: $T_0 + [95, 100]$ s | | C: $T_0 + [100, 107]$ s | | B + C: $T_0 + [95, 107]$ s | |
|---|---|---|---|---|---|---|
| | ISSM | ISSMExpCut | ISSM | ISSMExpCut | ISSM | ISSMExpCut |
| $\alpha$ | $-0.73 \pm 0.08$ | $-0.61 \pm 0.12$ | $-0.71 \pm 0.07$ | $-0.65 \pm 0.08$ | $-0.72 \pm 0.05$ | $-0.64 \pm 0.07$ |
| $\beta$ | $-2.50 \pm 0.06$ | $-2.10 \pm 0.09$ | $-2.47 \pm 0.04$ | $-2.28 \pm 0.07$ | $-2.48 \pm 0.04$ | $-2.21 \pm 0.07$ |
| $E_p$ [keV] | $680 \pm 70$ | $1900 \pm 1400$ | $760 \pm 60$ | $950 \pm 140$ | $730 \pm 50$ | $1100 \pm 230$ |
| $E_{\rm cut}$ [MeV] | ⋯ | **$34 \pm 11$** | ⋯ | $100 \pm 40$ | ⋯ | **$61 \pm 19$** |
| Norm. ($10^{-2}$) | $1.87 \pm 0.04$ | $1.87 \pm 0.04$ | $1.98 \pm 0.04$ | $1.99 \pm 0.04$ | $1.94 \pm 0.03$ | $1.94 \pm 0.03$ |
| $-\log(\mathcal{L})$ | 2128 | 2117 | 2349 | 2344 | 2788 | 2774 |
| $\sigma_{\rm cut}$ | ⋯ | **4.7** | ⋯ | 3.2 | ⋯ | **5.3** |

**Notes.** The units of the normalization are cm$^{-2}$ s$^{-1}$ keV$^{-1}$. The bold values correspond to the cases where the spectral model in the alternate hypothesis is preferred to the spectral model in the null hypothesis, as explained in Section 2.2.

Table 8
threeML Spectral Fits of Band with and without the Cutoff on GBM+LLE+LAT Data in Time Intervals B, C, and B + C

| Parameter | B: $T_0 + [95, 100]$ s | | C: $T_0 + [100, 107]$ s | | B + C: $T_0 + [95, 107]$ s | |
|---|---|---|---|---|---|---|
| | Band | BandExpCut | Band | BandExpCut | Band | BandExpCut |
| $\alpha$ | $-0.84 \pm 0.05$ | $-0.81 \pm 0.05$ | $-0.85 \pm 0.04$ | $-0.82 \pm 0.04$ | $-0.85 \pm 0.03$ | $-0.81 \pm 0.03$ |
| $\beta$ | $-2.34 \pm 0.04$ | $-1.95 \pm 0.07$ | $-2.34 \pm 0.03$ | $-2.08 \pm 0.05$ | $-2.34 \pm 0.02$ | $-2.04 \pm 0.04$ |
| $E_p$ [keV] | $380 \pm 40$ | $330 \pm 40$ | $440 \pm 40$ | $390 \pm 40$ | $411 \pm 28$ | $363 \pm 26$ |
| $E_{\rm cut}$ [MeV] | ⋯ | **$32 \pm 9$** | ⋯ | **$67 \pm 16$** | ⋯ | **$51 \pm 10$** |
| Norm. ($10^{-2}$) | $2.58 \pm 0.16$ | $2.74 \pm 0.20$ | $2.56 \pm 0.13$ | $2.71 \pm 0.15$ | $2.57 \pm 0.10$ | $2.73 \pm 0.12$ |
| $-\log(\mathcal{L})$ | 2149 | 2130 | 2388 | 2371 | 2840 | 2805 |
| $\sigma_{\rm cut}$ | ⋯ | **6.2** | ⋯ | **5.8** | ⋯ | **8.4** |

**Notes.** The units of the normalization are cm$^{-2}$ s$^{-1}$ keV$^{-1}$. The bold values correspond to the cases where the spectral model in the alternate hypothesis is preferred to the spectral model in the null hypothesis, as explained in Section 2.2.

Table 9
threeML Spectral Fits of ISSM with and without the Cutoff on GBM+LLE+LAT Data in Time Intervals B, C, and B + C

| Parameter | B: $T_0 + [95, 100]$ s | | C: $T_0 + [100, 107]$ s | | B + C: $T_0 + [95, 107]$ s | |
|---|---|---|---|---|---|---|
| | ISSM | ISSMExpCut | ISSM | ISSMExpCut | ISSM | ISSMExpCut |
| $\alpha$ | $-0.75 \pm 0.08$ | $-0.62 \pm 0.12$ | $-0.73 \pm 0.06$ | $-0.64 \pm 0.08$ | $-0.74 \pm 0.05$ | $-0.64 \pm 0.07$ |
| $\beta$ | $-2.53 \pm 0.06$ | $-2.12 \pm 0.09$ | $-2.51 \pm 0.04$ | $-2.27 \pm 0.07$ | $-2.52 \pm 0.04$ | $-2.22 \pm 0.05$ |
| $E_p$ [keV] | $670 \pm 70$ | $1500 \pm 900$ | $750 \pm 60$ | $980 \pm 140$ | $720 \pm 50$ | $1060 \pm 160$ |
| $E_{\rm cut}$ [MeV] | ⋯ | **$41 \pm 13$** | ⋯ | **$88 \pm 28$** | ⋯ | **$66 \pm 15$** |
| Norm. ($10^{-2}$) | $1.87 \pm 0.04$ | $1.87 \pm 0.04$ | $1.98 \pm 0.04$ | $1.99 \pm 0.04$ | $1.93 \pm 0.03$ | $1.94 \pm 0.03$ |
| $-\log(\mathcal{L})$ | 2147 | 2135 | 2383 | 2373 | 2832 | 2812 |
| $\sigma_{\rm cut}$ | ⋯ | **4.9** | ⋯ | **4.5** | ⋯ | **6.3** |

**Notes.** The units of the normalization are cm$^{-2}$ s$^{-1}$ keV$^{-1}$. The bold values correspond to the cases where the spectral model in the alternate hypothesis is preferred to the spectral model in the null hypothesis, as explained in Section 2.2.

Table 10
Summary of the threeML Spectral Analysis

| | | B: $T_0 + [95, 100]$ s | | C: $T_0 + [100, 107]$ s | | B + C: $T_0 + [95, 107]$ s | |
|---|---|---|---|---|---|---|---|
| | | GBM+LLE | GBM+LLE+LAT | GBM+LLE | GBM+LLE+LAT | GBM+LLE | GBM+LLE+LAT |
| BandExpCut | $E_{\rm cut}$ [MeV] | $23 \pm 6$ | $32 \pm 9$ | $71 \pm 25$ | $67 \pm 16$ | $43 \pm 11$ | $51 \pm 10$ |
| | $\sigma_{\rm cut}$ | 5.8 | 6.2 | 4.7 | 5.8 | 7.1 | 8.4 |
| ISSMExpCut | $E_{\rm cut}$ [MeV] | $34 \pm 11$ | $41 \pm 13$ | $100 \pm 40$ | $88 \pm 28$ | $61 \pm 19$ | $66 \pm 15$ |
| | $\sigma_{\rm cut}$ | 4.7 | 4.9 | 3.2 | 4.5 | 5.3 | 6.3 |

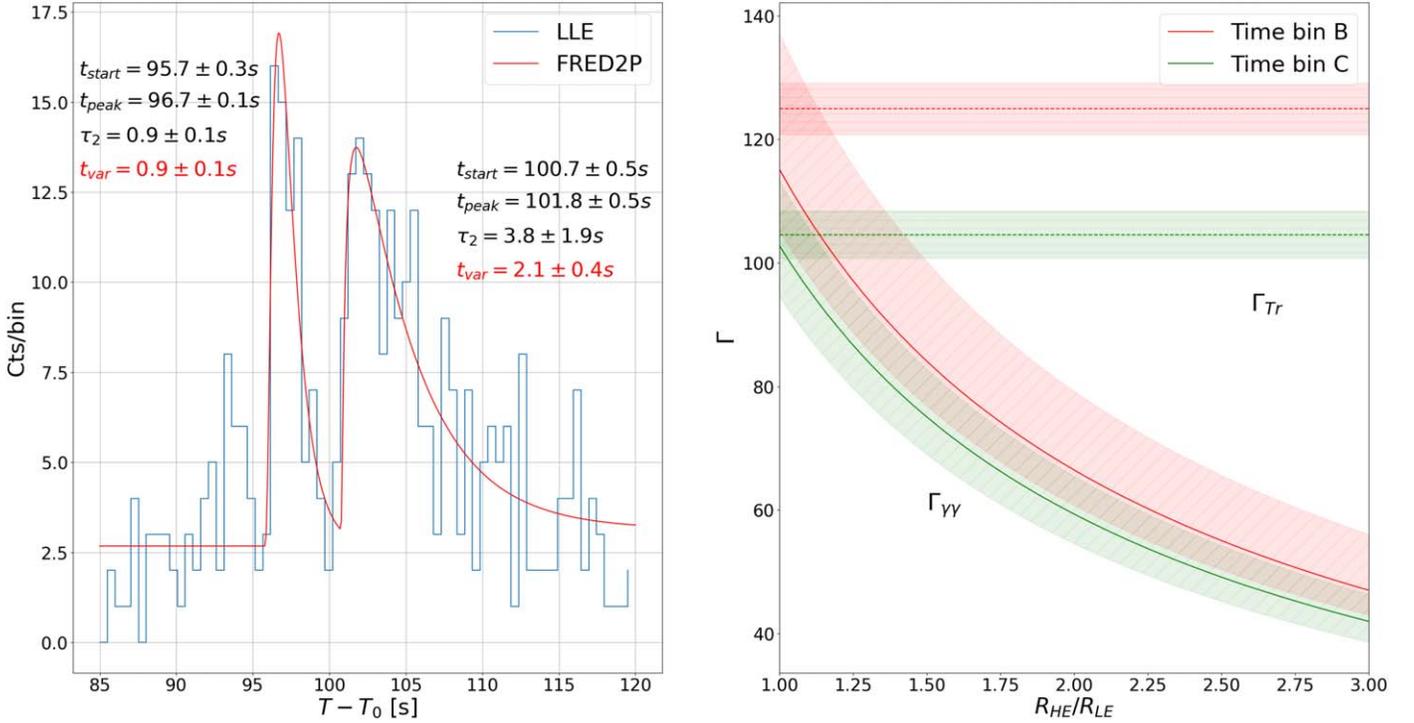

**Figure 4.** Left: light curve showing the two LLE peaks with the best-fit FRED2P function superimposed. Right: $\Gamma_{\gamma\gamma}$ and $\Gamma_{\text{Tr}}$ as a function of the ratio of the radii at which the high- and low-energy emissions were produced in time bins B and C.

uncertainty in our analysis. Specifically, we considered solid detections ($\sigma_{\text{cut}} > 4\sigma$), and we discarded the `pyXSPEC` ISSM fits that are slightly worse than the Band fits. We chose as final values the spectral cutoff energies obtained with the `pyXSPEC` fit of BandExpCut on GBM+LLE data, and we estimated the systematics from the absolute variations of the other analyses around these results, ignoring the statistical errors.

1. For time bin B, we consider the result of the `pyXSPEC` analysis $E_{\text{cut,Band}} = 23 \pm 8$ MeV in the opacity computation (see Section 4.2). The lowest value for such a cutoff is 23 MeV when fitting BandExpCut to GBM+LLE data with both `pyXSPEC` and `threeML`. The highest value is 41 MeV when fitting ISSMExpCut to GBM+LLE+LAT data. We thus estimate the cutoff value as $E_{\text{cut}} = 23 \pm 8$ (stat) $+18/-0$ (syst).
2. For time bin C, we consider $E_{\text{cut,Band}} = 69 \pm 20$ MeV. The lowest value of $E_{\text{cut}}$ is 67 MeV in the fit of BandExpCut to GBM+LLE+LAT data, and the highest value is 88 MeV in the fit of ISSMExpCut to the same data. We thus estimate the cutoff value as $E_{\text{cut}} = 69 \pm 20$ (stat) $+19/-2$ (syst).
3. In time bin B + C, $E_{\text{cut,Band}} = 42 \pm 11$ MeV. The lowest and highest values of $E_{\text{cut}}$ are 42 MeV in the `pyXSPEC` fit of BandExpCut and 66 MeV in the `threeML` fit of ISSMExpCut to GBM+LLE+LAT data. We thus estimate the cutoff value as $E_{\text{cut}} = 42 \pm 11$ (stat) $+24/-0$ (syst).

## 4. Interpretation

The temporal variability observed at high energy suggests that the detected spectral cutoffs are due to gamma opacity to an electron–positron pair creation. In the same spirit as Yassine et al. (2017) and the theoretical framework developed by Hascoët et al. (2012), we estimated the minimum variability timescale of the observed high-energy emission. Coupling the minimum variability timescale with the detected cutoffs, we determined the speed of the jet and localized the region in which all of the high-energy emission was produced.

### 4.1. Estimate of the Variability Timescale

In order to estimate the minimum variability timescale, we considered the fast rise exponential decay (FRED) function (Norris et al. 2005; Yassine et al. 2017), and we modified it to simultaneously fit the two main LLE observed peaks. This modified FRED function, which we call FRED2P, reads

$$I(t) = \begin{cases} B_x, & \text{if } t \leqslant t_{\text{start},x} \\ A_x \times \exp\left\{-\frac{1}{\tau_{2x}}\left[\frac{(t_{\text{peak},x} - t_{\text{start},x})^2}{t - t_{\text{start},x}} + (t - t_{\text{start},x})\right]\right\} + B_x, & \text{if } t_{\text{start},x} < t \leqslant t_{\text{start},y} \\ A_y \times \exp\left\{-\frac{1}{\tau_{2y}}\left[\frac{(t_{\text{peak},y} - t_{\text{start},y})^2}{t - t_{\text{start},y}} + (t - t_{\text{start},y})\right]\right\} + B_y, & \text{if } t > t_{\text{start},y} \end{cases}, \quad (3)$$

with

$$B_y = I(t_{start,y}). \quad (4)$$

The labels $x$ and $y$ refer to the first and second LLE peak, respectively. FRED2P is parameterized on each peak as the normalization A, offset B, start time of the pulse $t_{start}$, peak time of the pulse $t_{peak}$, and decay index $\tau_2$, which characterizes the decrease of the pulse. The left panel of Figure 4 shows the two LLE peaks superimposed on the best-fit FRED2P function. The minimum variability timescale of each pulse is estimated as the half-width at half-maximum and reads

$$t_{var} = \frac{\tau_2}{2} \times \sqrt{\left(\log(2) + 2\frac{t_{peak} - t_{start}}{\tau_2}\right)^2 - 4\left(\frac{t_{peak} - t_{start}}{\tau_2}\right)^2}. \quad (5)$$

The minimum variability timescale of the first peak is $t_{var,x} = 0.88 \pm 0.13$ s, and $t_{var,y} = 2.1 \pm 0.4$ s for the second peak.

### 4.2. Bulk Lorentz Factor and Localization of the Prompt Emission Region

The bulk Lorentz factor $\Gamma_{bulk}$ is obtained as in Yassine et al. (2017), assuming that the observed spectral cutoff is due to opacity to gamma–gamma annihilation in the GRB jet and that the prompt emission is produced near or above the photosphere at a radius $R_{LE}$ for the low-energy (MeV) emission and $R_{HE}$ for the high-energy (tens of MeV) emission. This opacity model was proposed by Hascoët et al. (2012) and applied by Yassine et al. (2017) to determine $\Gamma_{bulk}$ and the emission radii of GRB 090926A. The radius at which the low-energy emission is produced is obtained from the estimated variability as

$$R_{LE} = 2c\Gamma^2 \frac{t_{var}}{1+z}. \quad (6)$$

The $\Gamma_{bulk}$ of the jet is estimated directly as

$$\Gamma_{\gamma\gamma} = \frac{K\Phi(s)}{\left[\frac{1}{2}\left(1 + \frac{R_{HE}}{R_{LE}}\right)\left(\frac{R_{HE}}{R_{LE}}\right)\right]^{1/2}}(1+z)^{-(1+s)/(1-s)}$$
$$\times \left\{\sigma_T\left[\frac{D_L(z)}{ct_{var}}\right]^2 E_* F(E_*)\right\}^{1/2(1-s)}$$
$$\times \left[\frac{E_* E_{cut}}{(m_e c^2)^2}\right]^{(s+1)/2(s-1)}, \quad (7)$$

where $t_{var}$ is the estimated variability timescale in the considered time interval, $E_{cut}$ is the energy of the detected cutoff, $E_*$ is the typical energy of the photons interacting with those at the cutoff energy, $s$ is the photon index of the seed spectrum close to $E_*$, and $F(E_*)$ is the photon fluence at $E_*$ integrated over $t_{var}$. The values employed to compute $\Gamma_{\gamma\gamma}$ are reported in Table 11. In the error propagation, we also considered the systematic uncertainties of $E_{cut}$ reported at the end of Section 3.2, and we added them in quadrature to the statistical uncertainties.

The photospheric radius $R_{ph}$ at which the jet becomes transparent to Thomson scattering and the minimal bulk Lorentz factor $\Gamma_{Tr}$ defining this transparency condition are

**Table 11**
Summary of the Parameters Employed in the Computation of $\Gamma_{bulk}$ and the Observed Energy Emission Radius in Time Bins B, C, and B + C

| Time Bin | B: $T_0$ + [95, 100] s | C: $T_0$ + [100, 107] s | B + C: $T_0$ + [95, 107] s |
|---|---|---|---|
| $t_{var}$ [s] | 0.88 ± 0.13 | 2.1 ± 0.4 | 1.5 ± 0.5 |
| $s$ | −1.92 ± 0.10 | −2.11 ± 0.06 | −2.03 ± 0.05 |
| $\Phi$ (s) | 0.48 ± 0.01 | 0.47 ± 0.01 | 0.47 ± 0.01 |
| $E_{cut}$ [MeV] | 23 ± 8 | 69 ± 20 | 42 ± 11 |
| $E_*$ [MeV] | 1 | 1 | 1 |
| $F(E_*)$ [cm$^{-2}$ MeV$^{-1}$] | 0.34 ± 0.03 | 0.81 ± 0.04 | 0.57 ± 0.02 |
| $L$ [$10^{53}$ erg s$^{-1}$] | 7.6 ± 0.6 | 7.5 ± 0.3 | 7.6 ± 0.3 |
| $R_{LE}$ [$10^{14}$ cm] | 1.2 ± 0.3 | 2.4 ± 0.6 | 1.8 ± 0.7 |
| $\Gamma_{\gamma\gamma}$ ($R_{LE} = R_{HE}$) | 115 ± 10 | 103 ± 8 | 105 ± 13 |

**Note.** The luminosity is computed in the observer frame energy range 10 keV–1 GeV.

**Table 12**
Summary of the Radius at Which the Low-energy Emission Took Place $R_{LE}$, the Photospheric Radius $R_{ph}$, $\Gamma_{\gamma\gamma}$, and $\Gamma_{Tr}$ in Time Bins B, C, and B + C

| Time Bin | B: $T_0$ + [95, 100] s | C: $T_0$ + [100, 107] s | B + C: $T_0$ + [95, 107] s |
|---|---|---|---|
| $R_{LE}$ [$10^{14}$ cm] | 1.2 ± 0.3 | 2.4 ± 0.6 | 1.8 ± 0.7 |
| $R_{ph}$ [$10^{14}$ cm] | 1.9 ± 0.5 | 2.6 ± 0.6 | 2.4 ± 0.9 |
| $\Gamma_{\gamma\gamma}$ | 115 ± 10 | 103 ± 8 | 105 ± 13 |
| $\Gamma_{Tr}$ | 125 ± 4 | 105 ± 4 | 112 ± 8 |

computed as in Yassine et al. (2017),

$$R_{ph} \simeq \frac{\sigma_T \dot{E}}{8\pi c^3 m_p \bar{\Gamma}^3}, \quad (8)$$

where $\sigma_T = 6.65 \times 10^{-29}$ m$^2$ is the Thomson cross section, $\dot{E}$ is the total power injected in the flow, $m_p = 1.67 \times 10^{-27}$ kg is the proton mass, and $\bar{\Gamma} = \frac{1+\kappa}{2}\Gamma_{\gamma\gamma}$ is the average Lorentz factor in the flow, where $\kappa$ is the ratio between the highest and lowest values of $\Gamma_{bulk}$. The transparency condition $R_{LE} \geqslant R_{ph}$ translates to

$$\Gamma_{\gamma\gamma} > \Gamma_{Tr} \simeq \left[\frac{\sigma_T \dot{E}}{8\pi c^4 m_p t_{var}}\right]^{1/5}. \quad (9)$$

The values of the mentioned quantities are reported in Table 12. It is worth noting that the photospheric radii are of the order of $10^{14}$ cm, well above the typical range of $10^{10}$–$10^{11}$ cm. In fact, the high values for the luminosity and the moderate values of $\Gamma_{bulk}$ presented in Table 11 induce large photospheric radii, as shown by Equation (8).

The right panel of Figure 4 shows the value of $\Gamma_{\gamma\gamma}$ and $\Gamma_{Tr}$ as a function of the radii at which the high- and low-energy emissions were produced. The contours of $\Gamma_{\gamma\gamma}$ have been computed including the systematic errors estimated at the end of Section 3.2. We note that when the high- and low-energy emissions are cospatial, $\Gamma_{\gamma\gamma}$ and its contour are comparable to or greater than $\Gamma_{Tr}$ in time bins B and C. The transparency condition is thus fulfilled. We conclude that the bulk Lorentz factor of the jet in the prompt phase of GRB 220101A is $\Gamma_{bulk} \sim 110$ and that all of the high-energy emission took place near

**Table 13**
List of LAT-detected Bursts Presenting a Significant Cutoff at High Energies

| Burst | $z$ | $E_{\rm cut,obs}$ (MeV) | $E_{\rm cut,ref}$ (MeV) | $\Gamma_{\rm bulk}$ | Bibliography |
|---|---|---|---|---|---|
| GRB 090926A | 2.1062 | 370−50/+60 | 1150−155/+186 | 230−100 | Yassine et al. (2017) |
| GRB 100724B | Unknown | 20−60 | ⋯ | 100−400, depending on $z$ | Vianello et al. (2018) |
| GRB 160509A | 1.17 | 80−150 | 170−330 | 100−400 | Vianello et al. (2018) |
| GRB 170405A | 3.510 | 50 | 225 | 170−420 | Arimoto et al. (2020) |
| GRB 220101A | 4.618 | 40 | 230 | 105 | This work |

**Note.** For the ones with a redshift measurement, the cutoff energy is also reported in the source reference frame.

or above the photosphere at a radius of a few $10^{14}$ cm, typical of internal shocks.

### 5. Discussion and Conclusions

In this work, we assume that the observed variable emission is prompt emission. We note that Bianco et al. (2023) interpreted such early variable emission as afterglow and assumed that it is synchrotron radiation produced by a fast-spinning newborn neutron star that injects energy into the expanding supernova ejecta (Rueda et al. 2022). The authors considered the rest-frame temporal delay of the observed radiation and characterized the transition in the structure of the central neutron star in its first instants. Their analysis relies on the temporal delay of the radiation emitted, and it is an alternative to the interpretation we present in this paper.

The work of Moradi et al. (2021) on GRB 190114C pointed to a precise quantum electrodynamics model to explain the ultrarelativistic prompt emission of such a bright burst. The high energy budget of GRB 220101A makes it a companion to GRB 190114C, and a similar analysis would also be interesting in this case. However, the required detailed time-resolved analysis is beyond the scope of this work.

Table 13 lists the LAT-detected bursts that we found in the literature and presents a spectral cutoff at high energies. We stress that we did not consider the totality of the LAT-detected bursts, and we did not search systematically for the presence of an exponential spectral high-energy cutoff. This analysis shall be done in the future, and it is beyond the scope of this paper. For each of the mentioned bursts, we report the estimated values of $\Gamma_{\rm bulk}$, the spectral cutoff in the observer frame, and the spectral cutoff in the reference frame for the bursts with a redshift measurement. The value of $\Gamma_{\rm bulk}$ is 100–400. In the cases of GRB 090926A and GRB 220101A, $\Gamma_{\rm bulk}$ was determined following the procedure presented in the previous section. This estimation is based on the work of Hascoët et al. (2012), who accounted for the geometry of the GRB jet, and thus provides a realistic description of the jet dynamics. For GRB 100724B and GRB 160509A, Vianello et al. (2018) adopted two physical models: the semiphenomenological internal shock model developed by Granot et al. (2008), which provides a temporal, spatial, and directional dependence of the pair-production interaction and a conservative lower limit of $\Gamma_{\rm bulk}$, and the photospheric model by Gill & Thompson (2014). Vianello et al. (2018) estimated $\Gamma_{\rm bulk}$ in the interval 100–400 for these two bursts. In the case of GRB 170405A, Arimoto et al. (2020) estimated a lower limit of $\Gamma_{\rm bulk} = 170$ by applying the mentioned method of Granot et al. (2008) and provided an upper limit of $\Gamma_{\rm bulk} = 420$, which required the cutoff energy in the comoving frame to be $m_e c^2$: $\Gamma_{\rm bulk,max} = (1 + z)\frac{E_{\rm cut}}{m_e c^2}$ (Gill & Granot 2018).

In this work, we adopted the approach developed by Hascoët et al. (2012) and previously applied by Yassine et al. (2017) on GRB 090926A to directly estimate the bulk Lorentz factor and localize the region at which all of the high-energy emission of GRB 220101A took place. We stress that this approach does not rely on the specific emission process responsible for the detected emission and that the estimated $\Gamma_{\rm bulk}$ is comparable with the corresponding value of four other LAT-detected bursts that are well known for presenting a spectral cutoff at high energies. These bursts represent a precious set in which a direct estimation of $\Gamma_{\rm bulk}$ can be performed.

### Acknowledgments

The Fermi LAT Collaboration acknowledges generous ongoing support from a number of agencies and institutes that have supported both the development and the operation of the LAT, as well as scientific data analysis. These include the National Aeronautics and Space Administration and the Department of Energy in the United States; the Commissariat à l'Energie Atomique and the Centre National de la Recherche Scientifique/Institut National de Physique Nucléaire et de Physique des Particules in France; the Agenzia Spaziale Italiana and the Istituto Nazionale di Fisica Nucleare in Italy; the Ministry of Education, Culture, Sports, Science and Technology (MEXT), High Energy Accelerator Research Organization (KEK), and Japan Aerospace Exploration Agency (JAXA) in Japan; and the K. A. Wallenberg Foundation, the Swedish Research Council, and the Swedish National Space Board in Sweden.

Additional support for science analysis during the operations phase is gratefully acknowledged from the Istituto Nazionale di Astrofisica in Italy and the Centre National d'Études Spatiales in France. This work was performed in part under DOE contract DE-AC02-76SF00515.

### ORCID iDs

Lorenzo Scotton https://orcid.org/0000-0002-0602-0235
Frédéric Piron https://orcid.org/0000-0001-6885-7156
Nicola Omodei https://orcid.org/0000-0002-5448-7577
Niccolò Di Lalla https://orcid.org/0000-0002-7574-1298
Elisabetta Bissaldi https://orcid.org/0000-0001-9935-8106

### References

Abbott, B. P., Abbott, R., Abbott, T. D. & South Africa/MeerKAT, S. 2017, ApJL, 848, L12
Ackermann, M., Ajello, M., Asano, K., et al. 2013, ApJS, 209, 11
Arimoto, M., Asano, K., Tachibana, Y., & Axelsson, M. 2020, ApJ, 891, 106
Arimoto, M., Scotton, L., Longo, F. & Fermi-LAT Collaboration 2022, GCN, 31350, https://gcn.nasa.gov/circulars/31350

Arnaud, K. A. 1996, in ASP Conf. Ser. 101, Astronomical Data Analysis Software and Systems V, ed. G. H. Jacoby & J. Barnes (San Francisco, CA: ASP), 17
Atwood, W. B., Abdo, A. A., Ackermann, M., et al. 2009, ApJ, 697, 1071
Band, D., Matteson, J., Ford, L., et al. 1993, ApJ, 413, 281
Bianco, C. L., Mirtorabi, M. T., Moradi, R., et al. 2023, arXiv:2306.05855
Bloom, J. S., Kulkarni, S. R., & Djorgovski, S. G. 2002, AJ, 123, 1111
Bošnjak, Ž., & Daigne, F. 2014, A&A, 568, A45
de Ugarte Postigo, A., Kann, D. A., Thoene, C. C., et al. 2022, GCN, 31358, https://gcn.nasa.gov/circulars/31358?page=31
Eichler, D., Livio, M., Piran, T., & Schramm, D. N. 1989, Natur, 340, 126
Fu, S. Y., Zhu, Z. P., Xu, D., Liu, X., & Jiang, S. Q. 2022, GCN, 31353, https://gcn.nasa.gov/circulars/31353?page=30
Fynbo, J. P. U., de Ugarte Postigo, A., Xu, D., et al. 2022, GCN, 31359, https://gcn.nasa.gov/circulars/31359?page=33
Galama, T. J., Vreeswijk, P. M., van Paradijs, J., et al. 1998, Natur, 395, 670
Gehrels, N., Chincarini, G., Giommi, P., et al. 2004, ApJ, 611, 1005
Gill, R., & Granot, J. 2018, MNRAS, 475, L1
Gill, R., & Thompson, C. 2014, ApJ, 796, 81
Granot, J., Cohen-Tanugi, J., & Silva, E. D. C. E. 2008, ApJ, 677, 92
Hascoët, R., Daigne, F., Mochkovitch, R., & Vennin, V. 2012, MNRAS, 421, 525
Hentunen, V.-P., Nissinen, M., & Heikkinen, E. 2022, GCN, 31356, https://gcn.nasa.gov/circulars/31356
Hjorth, J., Sollerman, J., Møller, P., et al. 2003, Natur, 423, 847
Kouveliotou, C., Meegan, C. A., Fishman, G. J., et al. 1993, ApJL, 413, L101
Laskar, T. 2022, GCN, 31372, https://gcn.nasa.gov/circulars/31372?page=33
Lesage, S., Meegan, C. & Fermi Gamma-ray Burst Monitor Team 2022, GCN, 31360, https://gcn.nasa.gov/circulars/31360?page=31
Meegan, C., Lichti, G., Bhat, P. N., et al. 2009, ApJ, 702, 791
Mei, A., Oganesyan, G., Tsvetkova, A., et al. 2022, ApJ, 941, 82
Moradi, R., Rueda, J. A., Ruffini, R., et al. 2021, PhRvD, 104, 063043
Narayan, R., Paczynski, B., & Piran, T. 1992, ApJL, 395, L83
Neyman, J., & Pearson, E. S. 1928, Biometrika, 20A, 263
Norris, J. P., Bonnell, J. T., Kazanas, D., et al. 2005, ApJ, 627, 324
Paczynski, B. 1991, AcA, 41, 257
Pelassa, V., Preece, R., Piron, F., Omodei, N., & Guiriec, S. 2010, arXiv:1002.2617
Perley, D. A. 2022, GCN, 31357, https://gcn.nasa.gov/circulars/31357?page=30
Piran, T. 2004, RvMP, 76, 1143
Rueda, J. A., Li, L., Moradi, R., et al. 2022, ApJ, 939, 62
Tohuvavohu, A., Gropp, J. D., Kennea, J. A., et al. 2022, GCN, 31347, https://gcn.gsfc.nasa.gov/other/220101A.gcn3
Vianello, G., Gill, R., Granot, J., et al. 2018, ApJ, 864, 163
Vianello, G., Lauer, R. J., Younk, P., et al. 2015, arXiv:1507.08343
Wilks, S. S. 1938, Ann. Math. Statist., 9, 60
Woosley, S. E. 1993, ApJ, 405, 273
Yassine, M., Piron, F., Daigne, F., et al. 2020, A&A, 640, A91
Yassine, M., Piron, F., Mochkovitch, R., & Daigne, F. 2017, A&A, 606, A93